\documentclass[a4paper,11pt]{article}
\pdfoutput=1
\usepackage{jcappub} 

\usepackage{footnote}	
\usepackage{xcolor}
\usepackage{bigints}
\usepackage{multirow}
\usepackage{appendix}
\usepackage{comment}
\usepackage{hyperref}
\usepackage[normalem]{ulem}
\usepackage{bm,relsize}  
\usepackage{gensymb}
\usepackage{fancyhdr} 
\usepackage{amsfonts,amsmath,amssymb,amsthm,mathtools}

\renewcommand{\arraystretch}{1.2}

\graphicspath{{figs/}} 

\begin{document}
\thispagestyle{empty}
\vspace*{-2cm}  %
\hfill\footnotesize KCL-PH-TH/2025-19
\title{Dark Matter in X-rays: Revised XMM-Newton Limits and New Constraints from eROSITA}

\author[a]{Shyam Balaji,}
\author[a]{Damon Cleaver,}
\author[b,c]{Pedro De la Torre Luque,}
\author[d]{and Miltiadis Michailidis}

\affiliation[a]{Physics Department, King’s College London, Strand, London, WC2R 2LS, United Kingdom}
\affiliation[b]{Departamento de F\'{i}sica Te\'{o}rica, M-15, Universidad Aut\'{o}noma de Madrid, E-28049 Madrid, Spain}
\affiliation[c]{Instituto de F\'{i}sica Te\'{o}rica UAM-CSIC, Universidad Aut\'{o}noma de Madrid, C/ Nicol\'{a}s Cabrera, 13-15, 28049 Madrid, Spain}
\affiliation[d]{W. W. Hansen Experimental Physics Laboratory, Kavli Institute for Particle Astrophysics and Cosmology, Department of Physics and SLAC National Accelerator Laboratory, Stanford University, Stanford, CA 94305, USA}

\emailAdd{shyam.balaji@kcl.ac.uk}
\emailAdd{damon.cleaver@kcl.ac.uk}
\emailAdd{pedro.delatorre@uam.es}
\emailAdd{milmicha@stanford.edu}

\abstract{
We investigate two classes of dark matter (DM) candidates, sub-GeV particles and primordial black holes (PBHs), that can inject low-energy electrons and positrons into the Milky Way and leave observable signatures in the X-ray sky. In the case of sub-GeV DM, annihilation or decay into $e^+e^-$ contributes to the diffuse sea of cosmic-ray (CR) leptons, which can generate bremsstrahlung and inverse Compton (IC) emission on Galactic photon fields, producing a broad spectrum from X-rays to $\gamma$-rays detectable by instruments such as eROSITA and {\sc XMM-Newton}. For PBHs with masses below $\sim10^{17}$ g, Hawking evaporation similarly yields low-energy $e^\pm$, leading to comparable diffuse emission. Using the first data release from eROSITA and incorporating up-to-date CR propagation and diffusion parameters, we derive new constraints on both scenarios. For sub-GeV DM, we exclude thermally averaged annihilation cross sections in the range $\sim 10^{-27}$–$10^{-25} \ \mathrm{cm^3/s}$ and decay lifetimes of $\sim 10^{24}$–$10^{25}$ s for masses between 1 MeV and 1 GeV, with eROSITA outperforming previous X-ray constraints below $\sim 30$ MeV. For asteroid-mass PBHs, we set new bounds on the DM fraction based on their Hawking-induced emission. Finally, we revisit previous constraints in Ref.~\cite{Cirelli:2023tnx} and consequently Ref.~\cite{DelaTorreLuque:2023olp} derived from {\sc XMM-Newton}, finding that they were approximately four orders of magnitude too stringent due to the use of the instrument’s geometric solid angle rather than its exposure-weighted solid angle. Upon using the exposure-weighted solid angle, we show that the revised {\sc XMM-Newton} limits are slightly weaker than those from eROSITA. 
}

 \maketitle

\section{Introduction}

The nature of dark matter (DM) remains one of the most compelling open problems in modern physics. A wealth of astrophysical and cosmological observations indicate that approximately 85\% of the matter content of the Universe is composed of DM~\cite{ParticleDataGroup:2024cfk, Cirelli:2024ssz}. Yet, its fundamental properties continue to elude direct detection. While weakly interacting massive particles (WIMPs) have long been leading candidates, their persistent non-observation~\cite{Cirelli:2024ssz} has intensified the exploration of alternative scenarios, including sub-GeV DM~\cite{Feng:2008ya, Hochberg:2014dra, Petraki:2013wwa, DelaTorreLuque:2024fcc} and primordial black holes (PBHs)~\cite{Carr:2020gox, Auffinger:2022khh, Escriva:2022duf}.

Sub-GeV DM candidates, with masses in the MeV–GeV range, are theoretically well-motivated and arise in a variety of extensions of the Standard Model~\cite{Petraki:2013wwa, Cirelli:2020bpc, Cirelli:2024ssz, Aghaie:2025dgl}. However, their low mass poses detection challenges. Direct detection experiments lose sensitivity in this regime, while indirect searches are hindered by charged particle suppression in the heliosphere and the limited sensitivity of existing $\gamma$-ray instruments in the keV–GeV band, the so-called “MeV gap”~\cite{De_Angelis_2021, Cirelli:2020bpc, Boudaud:2016mos}. Nevertheless, DM annihilation or decay into $e^\pm$ can inject low-energy electrons and positrons into the interstellar medium. These secondaries interact with Galactic radiation fields via bremsstrahlung radiation and inverse Compton (IC) scattering, producing X-rays in the keV to MeV range~\cite{Cirelli:2023tnx}, which offers a valuable observational window.

PBHs, originally proposed decades ago~\cite{Zeldovich:1967lct, Hawking:1971ei}, have reemerged as viable DM candidates, particularly in light of null results from particle-based searches. PBHs with masses $M_{\rm BH} \lesssim 10^{17}$ g can emit high-energy particles via Hawking evaporation~\cite{1974Natur.248...30H, MacGibbon:1991vc, Auffinger:2022khh}. Those below $10^{16}$ g can emit charged particles that generate a diffuse X-ray signal, similarly to sub-GeV DM scenarios. While these low-energy particles are difficult to detect directly on Earth due to solar modulation, observations from {\sc Voyager 1} beyond the heliopause have enabled constraints to be derived~\cite{cummings2016galactic, Boudaud:2018hqb}.

X-ray observations thus offer a uniquely complementary approach to probing both sub-GeV DM and PBHs. They are particularly sensitive to models involving low-energy $e^\pm$ injection and probe the MeV gap inaccessible to most $\gamma$-ray telescopes. Past studies using instruments like {\sc XMM-Newton} have searched for diffuse X-ray signals from light DM~\cite{Cirelli:2023tnx} and PBHs~\cite{DelaTorreLuque:2024qms}. Despite the long exposure times achievable with pointing telescopes, their narrow fields of view limit coverage to specific sky regions, constraining their ability to study large-scale emission or high-latitude backgrounds.

The Extended ROentgen Survey with an Imaging Telescope Array (eROSITA) represents a major advance for such searches. Launched aboard the Spectrum-Roentgen-Gamma (SRG) satellite in 2019~\cite{2021A&A...656A.132S}, eROSITA provides the most sensitive full-sky coverage in the 0.2–8 keV range, including the first complete view of the X-ray sky above 2 keV~\cite{2021A&A...647A...1P}. Even with only half its planned surveys completed, eROSITA already surpasses ROSAT in soft X-ray sensitivity, owing to its order-of-magnitude larger effective area above 0.3 keV and its degree-scale field of view. These features, including large sky coverage, high angular and spectral resolution, and deep exposure, make eROSITA ideally suited to search for both secondary X-ray emission from DM-induced $e^\pm$, and for Hawking radiation signatures from PBH evaporation across the Galactic halo.

While the Galactic Centre (GC) offers enhanced DM densities and signal brightness, eROSITA’s full-sky reach allows constraints to be derived from high-latitude regions where astrophysical backgrounds are lower. Prior eROSITA analyses have targeted sterile neutrinos and axion-like particles using the eFEDS field~\cite{Dekker:2021bos, Barinov:2022kfp, Fong:2024qeq}, but no comprehensive study has yet leveraged eROSITA data to constrain sub-GeV DM or PBHs with full consideration of CR propagation and secondary emission.

In this work, we address this gap by placing new constraints on light DM and sub-asteroid mass PBHs using eROSITA’s full-sky X-ray maps. We model continuum emission from DM-induced low-energy $e^\pm$ and Hawking evaporation from PBHs, deriving upper limits on the annihilation cross section, decay lifetime, and PBH DM fraction across a wide mass range. We then compare these constraints with existing bounds from other probes, demonstrating the complementarity and reach afforded by eROSITA. Additionally, we revisit constraints in Refs.~\cite{Cirelli:2023tnx,DelaTorreLuque:2023olp} derived from {\sc XMM-Newton} data~\cite{XMM}, showing that applying the geometric solid angle instead of the exposure-weighted solid angle led to flux differences of approximately four orders of magnitude. After addressing this, the {\sc XMM-Newton} limits are found to be weaker than those from eROSITA, realigning the landscape of X-ray DM constraints. We note that the above discussion does not apply to Ref.~\cite{XMM} which derives constraints using line-like signals from decaying DM from {\sc XMM-Newton}.

This paper is structured as follows. In Section~\ref{sec:eROSITA}, we describe the eROSITA dataset. Section~\ref{sec:setup} details our signal modelling and data analysis framework. In Section~\ref{sec:results}, we present our results and compare them with existing constraints. Section~\ref{sec:XMM} discusses the reanalysis of {\sc XMM-Newton} data and the impact of correcting the exposure estimate. Finally, we summarise our findings and conclude in Section~\ref{sec:Conclusion}.

\section{eROSITA diffuse emission - Data analysis}
\label{sec:eROSITA}

\begin{table}[t!]
\renewcommand{\arraystretch}{1.5}
\setlength{\tabcolsep}{4pt} 
\centering
\caption{Comparison of eROSITA with other X-ray satellites covering the keV band.}
\begin{tabular}{@{}llllll@{}}
\hline\hline
Telescope & Energy range & Effective area & FoV & Spatial res. & Sensitivity \\
 & (keV) & (cm$^2$ at 1 keV) & (\degree) & (arcsec) & (erg cm$^{-2}$ s$^{-1}$) \\ \hline
eROSITA & 0.2-10 & 1237 & 1 & $<$10 & $\sim$5$\times$10$^{-14}$ \\
 & & & & & (eRASS1, $<$2.3 keV) \\
{\sc {\sc XMM-Newton}} & 0.15-12 & $\sim$ 1200 (250) & 0.5 & 6 & $\sim$10$^{-14}$ \\
 & & total (single MOS) & & & (after 10 ks) \\
Chandra & 0.1-10 & 555 & 0.5 & 0.5 & 4$\times$10$^{-15}$ \\
 & & & & & (after 100 ks) \\
ROSAT & 0.1-2.4 & 400 & 2 & $\sim$30 & 1.5$\times$10$^{-13}$ \\ \hline
\end{tabular}
\label{tab}
\end{table}

This study is motivated by the technological advancements introduced by the recently launched eROSITA telescope in the X-ray domain. eROSITA comprises seven parallel-aligned X-ray telescope modules equipped with single chip-PN CCD detectors, which are similar to but improved upon those of {\sc {\sc XMM-Newton}} PN (discussed below). Each module features a $1\degree$ field of view (FoV) and individually hosts 54 nested mirror shells \cite{2021A&A...647A...1P}. This configuration offers our most sensitive view of the X-ray sky.

Although eROSITA is not a pointed X-ray instrument and therefore does not broadly permit X-ray observations lasting tens or hundreds of kiloseconds, it is significant to demonstrate its major advancements compared to other X-ray instruments operating in the same energy range (or there is at least a strong overlap) of $\sim0.2$-$10$~keV, such as ROSAT (the predecessor of the eROSITA all-sky survey), {\sc {\sc XMM-Newton}} and Chandra. 

The combined effective area from the seven eROSITA telescope modules is substantially greater than that of ROSAT and Chandra, though only slightly larger compared to {\sc {\sc XMM-Newton}} (MOS+PN). However, what distinguishes eROSITA is its enhanced spectral resolution compared to {\sc {\sc XMM-Newton}} PN and the combination of the large effective area with the $1\degree$ FoV, which introduces the so-called grasp concept (the product of the effective area and FoV), which is significantly higher (by many times) than that of any current X-ray survey in the 0.3-3.5 keV energy range, as illustrated in Ref.~\cite{2021A&A...647A...1P} (Fig.~10). For a direct comparison, we summarise the aforementioned instruments' capabilities in Tab.~\ref{tab}.

Although the initial objective was to conduct eight all-sky surveys over a four-year period, only four surveys have been successfully completed to date. The corresponding datasets are not publicly accessible in their entirety. As such, this research utilises the newly available eRASS1 X-ray all-sky survey datasets, which became publicly available in February 2024, and are characterised by unprecedented sensitivity. The aim is to exploit the soft broadband eROSITA maps (0.3-2.3 keV) for the purposes of this DM search. 

The calibrated eRASS1 data were processed using the eROSITA Standard Analysis Software (eSASS) \cite{2022A&A...661A...1B}, adhering to all standard recommendations for data reduction from the eROSITA collaboration, which include exposure correction and the exclusion of bad time intervals and flaring events. Since the eROSITA survey consists of 4700 overlapping distinct sky tiles, of $3.6\degree\times3.6\degree$ each, the map reported in this study was produced by assembling the exposure corrected count images of the adjacent sky tiles that cover the selected region to be studied. 

The maps are produced in HEALPix format, using the healpy package \cite{2019JOSS....4.1298Z}, with a pixel size of $0.22\degree\times0.22\degree$ in units cts/s/$\mathrm{deg}^2$ (and subsequently converted to keV/sr/$\mathrm{cm^2}$/s for the purposes of the analysis) after properly filtering out point sources (according to the eRASS1 source catalog \cite{2024A&A...682A..34M}) above a selected flux threshold (point sources with \texttt{$\textrm{ML\_FLUX\_P2+ML\_FL}$}
\texttt{$\textrm{UX\_P3}$} in the eRASS1 source catalog above $10^{-13}$ were removed) and removing the particle background. This threshold is selected  such that 19\% of the cosmic X-ray background is removed, as discussed in Fig.~12 of the main data release collaboration paper~\cite{2024A&A...682A..34M}. 
Overall, the individual steps outlined by the eROSITA collaboration in Section 2 of Ref.~\cite{2024A&A...681A..77Z} were adapted to obtain the corresponding diffuse X-ray map from the selected region.

From this analysis, we obtain a full-sky map of the unresolved-diffuse emission observed by eROSITA in the $\sim 0.1$-$2.3$~keV band. Given the angular resolution of the telescope and its poor sensitivity for source detection (poor, at least, in comparison to {\sc XMM-Newton} or Chandra), this emission results from the sum of the truly diffuse emission, originated from the IC emission of CR electrons off the interstellar radiation fields (ISRFs)~\cite{DelaTorreLuque:2023olp} and the unresolved sources below the detection threshold of eROSITA. Therefore, with more observation time it will be possible to lower the diffuse background observed by eROSITA and improve current constraints.

\begin{figure}
    \centering
    \begin{minipage}{0.49\linewidth}
        \centering
        \includegraphics[width=\linewidth]{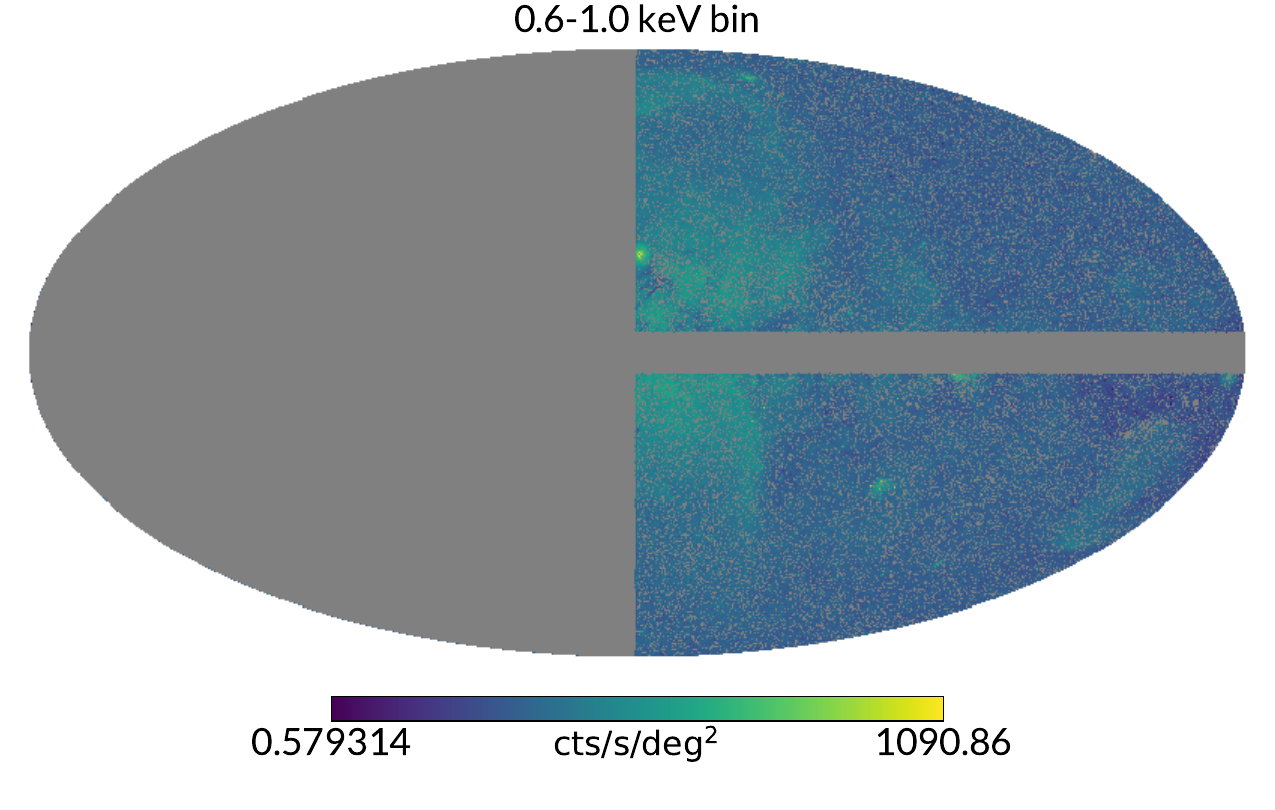}
    \end{minipage}
    \hfill
    \begin{minipage}{0.49\linewidth}
        \centering
        \includegraphics[width=\linewidth]{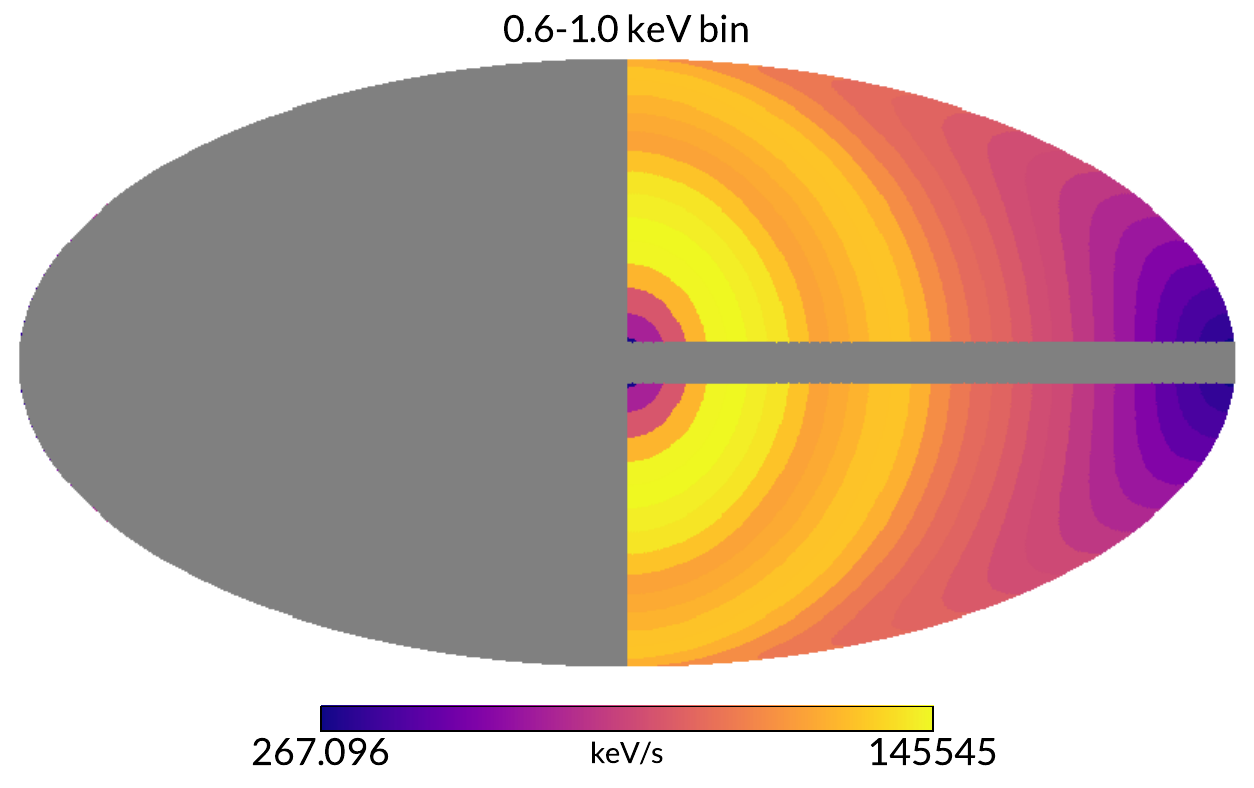}
    \end{minipage}
    \caption{Mollweide projections of the X-ray sky observed by eROSITA. The left panel shows the map with the Galactic plane mask applied ($|b| < 5^\circ$), while the right panel displays the concentric $6^\circ$ ringed regions used for subdivision of the Galaxy that will be used to derive constraints on dark matter, indicating the observed count rate as a color bar.}
    \label{fig:mollewides}
\end{figure}

Subsequently, we divide the Galaxy into 30 sub-regions that consist of $6^\circ$ wide concentric rings around the GC, following the strategy followed by Ref.~\cite{XMM}, with the {\sc XMM-Newton} data. This allows a consistent comparison between the eROSITA and the {\sc XMM-Newton} diffuse emission in the $0.1$ to $\sim10$~keV range, that already reveals that previous {\sc XMM-Newton} constraints have been overestimated. However, since the X-ray emission absorption by interstellar gas becomes more important at lower energy, we ignore the region of latitude $|b| < 5^\circ$ to mask the Galactic plane, as opposed to the analysis {\sc XMM-Newton} data, which used $|b| < 2^\circ$. Therefore, our result leads to more conservative constraints. We show an example of Mollewide projections of the eROSITA HEALPix data and the rings used in Fig.~\ref{fig:mollewides}. 

\section{Continuum signals from dark matter}
\label{sec:setup}

\subsection{Sub-GeV dark matter annihilation and decay }
\label{sec:subGeVDMannihilationanddecay}
In this work we implement the injection and propagation of $e^\pm$ from light DM, using a modified version of the {\tt DRAGON2} code~\citep{DRAGON2-1, DRAGON2-2}, which solves the CR diffusion-advection-loss equation~\citep{Ginz&Syr}. The DM injection spectra are taken from Ref.~\cite{Cirelli:2020bpc}, where we refer the reader for a comprehensive discussion of their production spectra.  For annihilation or decay, the crucial input of interest is the injection rate of electrons and positrons per unit volume, which is denoted $Q_e(\vec{x},E)$ and takes the form
\begin{equation}
    Q_e = 
    \begin{cases}
        \frac{\langle \sigma v\rangle}{2} \left(\frac{\rho_\chi(\vec{x})}{m_\chi}\right)^2 \frac{dN_e}{dE_e} & \text{(annihilation)} \\
        \Gamma \left(\frac{\rho_\chi(\vec{x})}{m_\chi}\right) \frac{dN_e}{dE_e} & \text{(decay)},
    \end{cases}
\label{eq:Source}
\end{equation}
with $\langle \sigma v \rangle$, $\Gamma$, $\rho_\chi$ and $\frac{dN_e}{dE_e}$ denoting the annihilation cross section, decay rate, DM density profile and spectrum, respectively. The above formulae assumes Majorana DM particles, an additional factor of 1/2 has to be included for annihilation in the case of Dirac fermionic DM since $\langle\sigma v\rangle$ has to be averaged 
 over particles and antiparticles. In this work, we will proceed assuming Majorana DM as our base case, which can easily be rescaled in the case of Dirac DM.

We restrict our analysis to the channel of direct  $ e^+e^-$ production, from a minimum kinematically allowable mass of $m_e$ ($2m_e$) for annihilation (decay) respectively. In both cases, we take the maximum mass of DM to be $1000$~MeV. We use a 2D axisymmetric Galactic model with a NFW DM profile~\citep{Navarro:1995iw} representing our base case.  Other annihilation (decay) channels such as $\mu^+ \mu^-$ and $\pi^+ \pi^-$ can also be considered. While these yield softer secondary $e^\pm$ spectra than direct $e^+e^-$, their production thresholds require higher DM masses. The resulting $e^\pm$ are therefore injected at higher absolute energies, in a regime where constraints from instruments such as INTEGRAL already dominate, making the eROSITA bounds comparatively weaker~\cite{Cirelli:2023tnx,DelaTorreLuque:2023olp}.
The propagation parameters used in this work are obtained from a fit to AMS-02 data~\cite{DeLaTorreLuque:2021ddh, Luque:2021nxb}, and are listed in Table~I of Ref.~\cite{DelaTorreLuque:2023olp}, where more of the technical details are given. The interplay between diffusion and energy losses is key at sub-GeV energies. Energy losses in the {\tt DRAGON2} code include ionisation, Coulomb, bremsstrahlung, IC, synchrotron and in-flight positron annihilation processes. The spatial and energy resolution employed is $\sim150$~pc and $5$\%.

The resulting $e^\pm$ distributions are passed to the {\tt HERMES} code~\citep{Dundovic:2021ryb} to compute IC X-ray emission, integrating over the ISRFs~\citep{Vernetto:2016alq}. 

Although IC losses have little effect on the $e^\pm$ spectra, the emission dominates the X-ray signal, especially below the MeV range that is relevant for experiments such as eROSITA and {\sc XMM-Newton}. We estimate the ISRF modelling uncertainty to be below 30\%~\citep{Vernetto2016prd}, and we separately account for final-state radiation (FSR) from the outgoing $e^\pm$, see Eq.~(6) of Ref.~\cite{Cirelli:2020bpc}. This contribution is sometimes referred to as “internal bremsstrahlung” in the DM literature (eg. Ref.~\cite{Bringmann_2012}). We emphasise that we do not include additional model-dependent internal bremsstrahlung from intermediate states, which depends on the underlying particle physics model. Full simulation inputs and outputs are available in Ref.~\citep{de_la_torre_luque_2023_10076728}.

\begin{figure}
    \centering
    \includegraphics[width=0.5\linewidth]{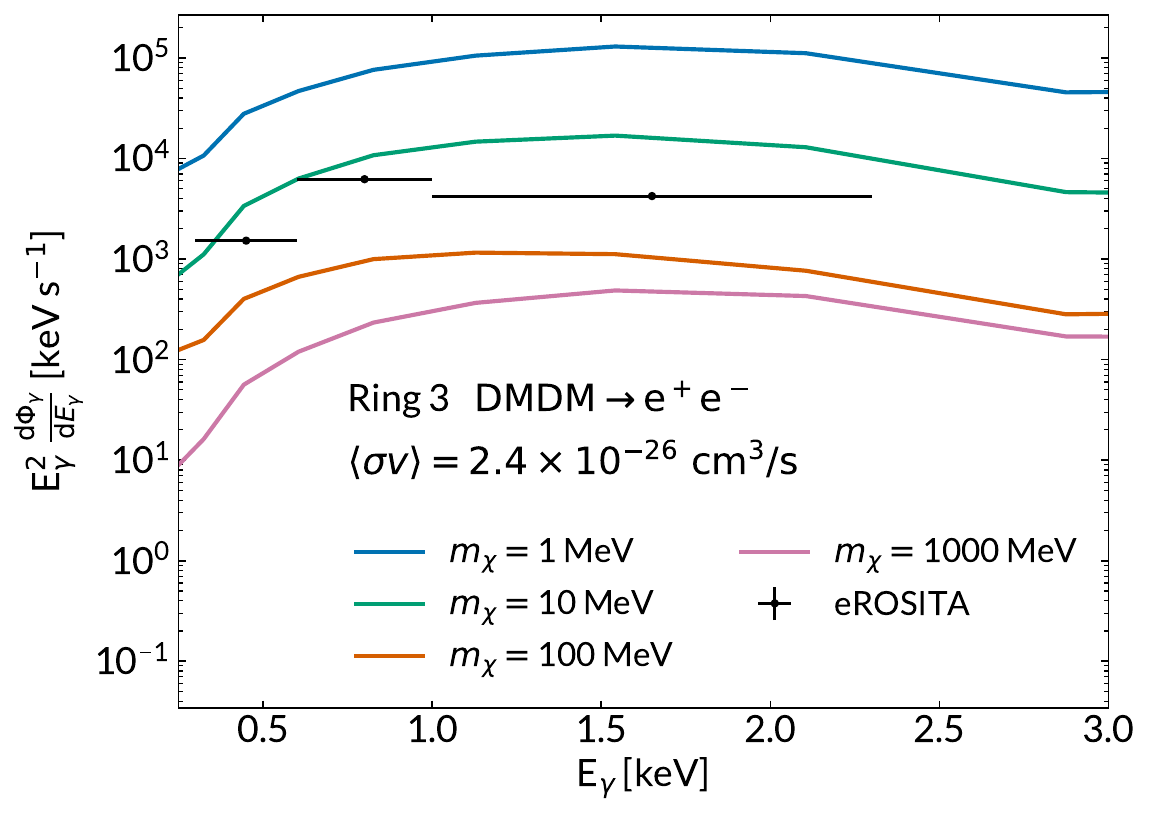}
\includegraphics[width=0.5\linewidth]{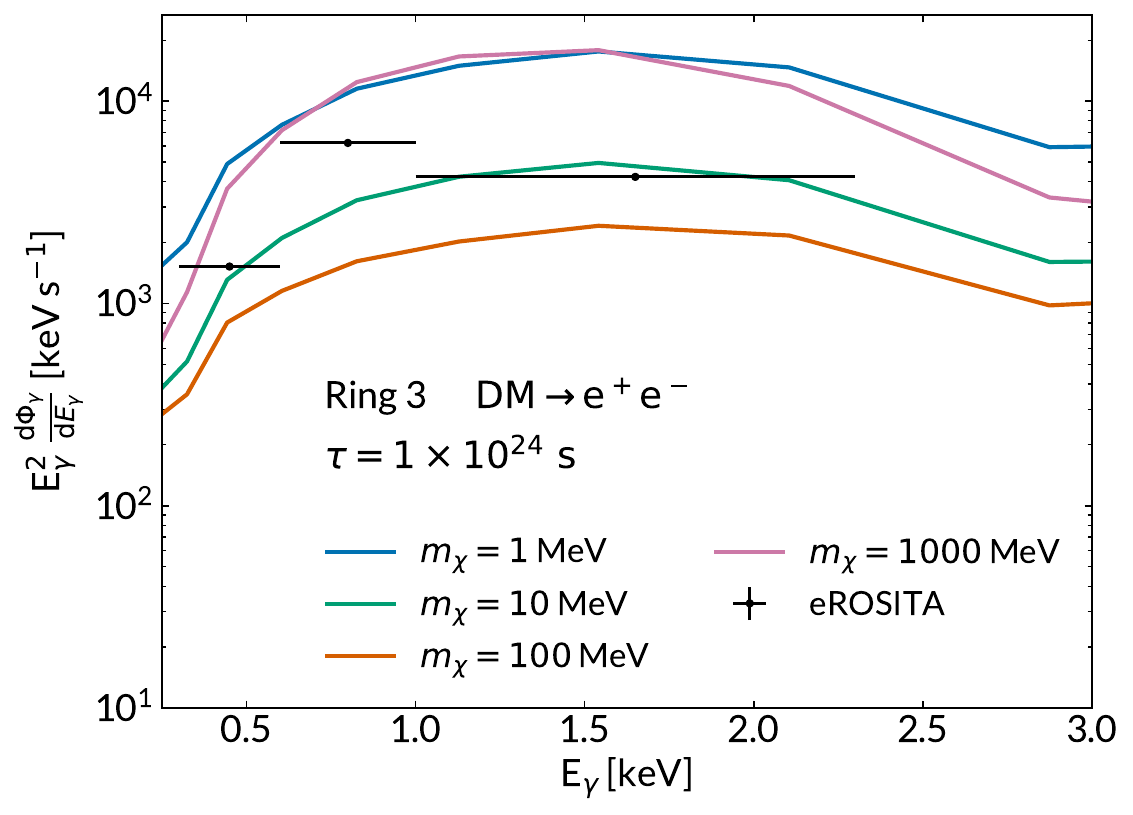}
\includegraphics[width=0.5\linewidth]{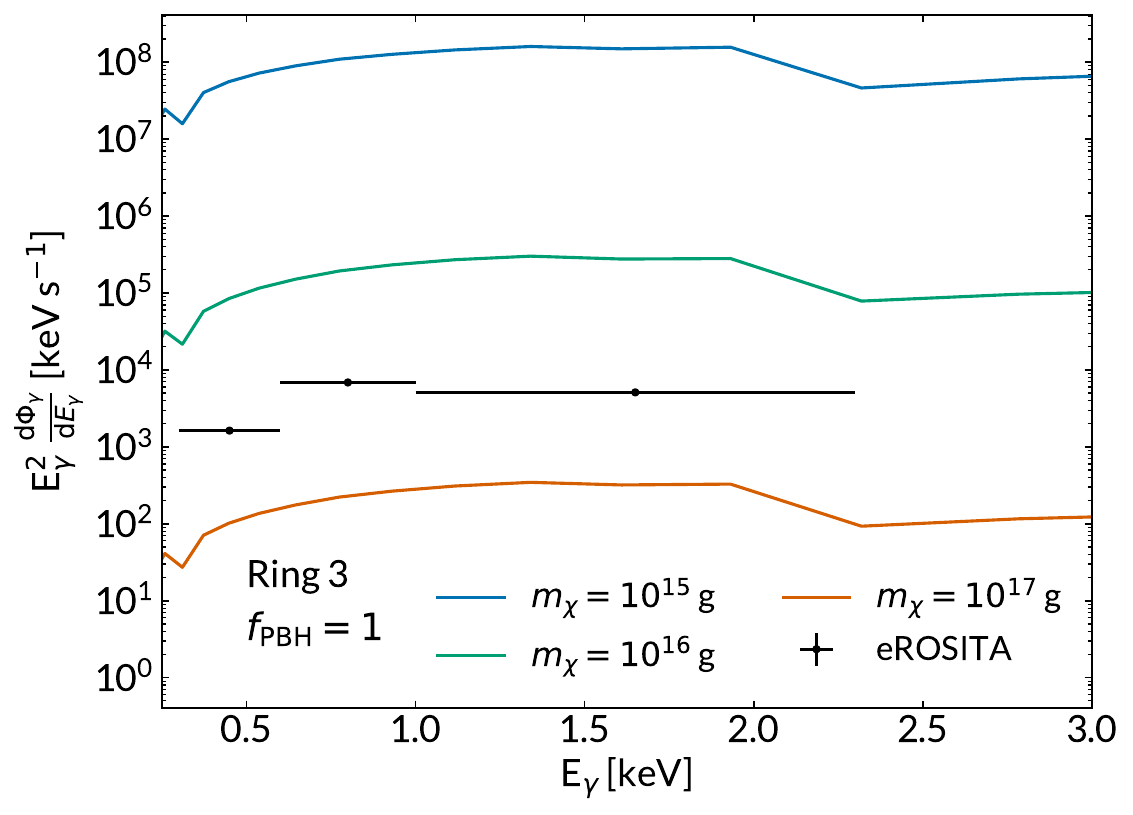}
    \caption{Comparison of eROSITA data (black points) in Ring 3 to the predicted particle DM or PBH evaporation induced X-ray signal for annihilation with $\langle\sigma v\rangle = 2.4 \times 10^{-26} \ \mathrm{cm^3/s}$ (top left) and decay with $\tau = 1 \times 10^{24} \ \mathrm{s}$ (top right) and for PBHs with $f_{\mathrm{PBH}} = 1$ (bottom). We show the predicted flux for different particle DM and PBH masses, as specified in the legends.}
    \label{fig:flux_plots}
\end{figure}

In Fig.~\ref{fig:flux_plots} we show examples of the expected X-ray flux spectra in eROSITA in the ring 3 region of the Galaxy for DM masses $m_\chi$ of 1, 10, 100, 1000 $\mathrm{MeV}$ for both annihilation and decay, with fixed $\langle\sigma v\rangle = 2.4 \times 10^{-26} \ \mathrm{cm^3/s}$ and $\tau = 1 \times 10^{24} \ \mathrm{s}$ respectively (colored lines). We show the measured flux of eROSITA for comparison (black points).

On top of the IC emission, we also consider the contribution from FSR for every mass, which is computed following Ref.~\cite{Cirelli:2023tnx}. This contribution becomes more relevant for lower masses, being important only below $\sim 10$~MeV. In our case, this contribution remains negligible when taking into account reacceleration, as shown in Ref.~\cite{DelaTorreLuque:2023olp}.

\subsection{Primordial black hole evaporation}
\label{sec:pbhevaporation}
Low mass black holes are theoretically expected to evaporate via Hawking radiation, emitting particles with energies up to the black hole temperature $T$~\cite{Hawking:1975vcx, Arbey:2019jmj}, which depends on its mass $M$, spin parameter $a \equiv J/M$ and charge
\begin{equation}
    T = \frac{1}{2\pi} \left(\frac{r_+ - M}{r_+^2 + a^2}\right), \quad r_+ = M + \sqrt{M^2 - a^2}.
\end{equation}
For Schwarzschild black holes ($a=0$), this reduces to $T = 1/(8\pi M)$. The primary particle spectrum for species index $i$ is then
\begin{equation}
    \frac{d^2N_i}{dtdE_i} = \frac{1}{2\pi} \sum_\text{d.o.f.} \frac{\Gamma_i(E_i,M,a^\star)}{e^{E'_i/T}\pm 1},
\end{equation}
where the energy $E'_i \equiv E_i - m\Omega$, $\Omega = a^\star/(2r_+)$, and $\Gamma_i$ are greybody factors accounting for gravitational redshift. We compute these spectra using \texttt{BlackHawk v2.2}~\cite{Arbey:2019mbc, Arbey:2021mbl}. 

To model sub-GeV $e^\pm$ relevant for X-ray production, we use \texttt{Hazma}~\cite{Coogan:2019qpu} for secondary particle spectra in the domain $M \sim 10^{14.5}$–$10^{17.5}$ g, where evaporation produces $e^\pm$, $\gamma$, and $\nu_{e,\mu,\tau}$. This accounts for decays of $\mu^\pm$, $\pi^{0,\pm}$ and FSR.

The injection rate of $e^\pm$ from PBHs per unit volume is
\begin{equation}
    Q_e(E_e,\vec{x}) = f_\textrm{PBH}\rho_\textrm{DM}(\vec{x}) \int_{M_\textrm{min}}^\infty \frac{dM}{M}\frac{dN_\textrm{PBH}}{dM} \frac{d^2N_e}{dtdE_e},
\end{equation}
where $f_\textrm{PBH}$ is the PBH fraction of DM, and $M_\textrm{min} \approx 7.5 \times 10^{14}$ g is the minimum mass of PBHs surviving today. For simplicity we adopt a monochromatic mass distribution ($\frac{dN_\textrm{PBH}}{dM}=\delta(M-M_\textrm{PBH})$) and assume non-rotating black holes, which constitute the most conservative scenario.

\subsection{Dark matter density profiles}
\label{sec:Densities}

In this analysis, we investigate three widely studied density profiles for describing the distribution of DM: the Navarro-Frenk-White (NFW) profile \cite{Navarro:1995iw,Navarro:1996gj} as our base case, a contracted Navarro-Frenk-White (cNFW) profile for a more cuspy scenario, which accounts for baryonic contraction \cite{Blumenthal:1985qy,Gnedin:2004cx} and the isothermal profile \cite{Begeman:1991iy,Bahcall:1980fb} for a more cored conservative scenario. These three benchmarks are useful to understand the uncertainties in the constraints derived from the assumption of the DM density distribution.

The NFW profile, derived from numerical N-body simulations of cold DM halos, characterises the DM density as
\begin{equation}
\rho_{\mathrm{NFW}}(r) = \frac{\rho_s}{\frac{r}{r_s}\left(1 + \frac{r}{r_s}\right)^2},
\end{equation}
where \( \rho_s \) is the characteristic density, set from the assumption that the local DM density is $0.4$~GeV/$\mathrm{cm^3}$ and \( r_s = 20~\text{kpc} \) \cite{Cirelli:2010xx}. 

However, baryonic processes, such as gas cooling and subsequent star formation, are expected to contract the DM halo inwards, leading to higher central densities. To account for this baryonic contraction, we adopt the cNFW profile. This modified profile uses the standard adiabatic contraction formalism introduced by Ref.~\cite{Blumenthal:1985qy} and refined by Ref.~\cite{Gnedin:2004cx}. The cNFW profile can significantly enhance the central DM density, thus strengthening observational signatures and constraints. The contracted NFW profile can be written
\begin{align}
    \rho_{\mathrm{cNFW}}(r) = \frac{\rho_s}{\left(\frac{r}{r_s}\right)^{\gamma}\left(1 + \frac{r}{r_s}\right)^{3-\gamma}},
\end{align}
where the normalisation \( \rho_s \) is again given by fixing the local DM density to $0.4$~GeV/$\mathrm{cm^3}$, the scale radius is the same as for the NFW case and we set the inner slope to $\gamma=1.4$.

Additionally, we consider an isothermal profile as a pessimistic scenario. The isothermal DM profile is a spherically symmetric, cored density profile that assumes DM behaves like an ideal gas in thermal equilibrium, leading to a flat central density core. The expression for the density profile can be written
\begin{equation}
\rho_\textrm{iso}(r) = \frac{\rho_0}{1 + \left(\frac{r}{r_c}\right)^2},
\end{equation}
with a typical scale radius of $r_c=3.5$~kpc~\cite{DRAGON2-1, Cirelli:2010xx}. This profile is often employed to explore scenarios with minimal astrophysical signals arising from DM annihilation and decay, thereby providing conservative limits on DM properties.

In subsequent sections, we consider the NFW profile as our base case, but also apply the cNFW and isothermal profiles when parameterising DM profile uncertainties, thereby evaluating the impact of uncertainty in the DM concentration in the centre of the Galaxy when evaluating DM indirect detection signals.

\section{Constraints from eROSITA data}
\label{sec:results}

Following the strategy from previous works such as  Refs.~\cite{Cirelli:2023tnx, DelaTorreLuque:2023olp}, to derive constraints on the DM thermally averaged cross-section $\langle\sigma v\rangle$ and lifetime $\tau$ for annihilation and decay into $e^+e^-$ respectively, as well as the fractional abundance $f_\mathrm{PBH} = \rho_{\mathrm{PBH}}/\rho_{\mathrm{DM}}$ in the case of PBHs, we use a combined analysis of all rings by using the chi-squared statistic, 
\begin{equation}
    \chi^2 = \sum_i \left(\frac{\mathrm{Max}[\phi_{\mathrm{DM},i}(p, m_\chi) -\phi_{\mathrm{eROSITA},i},\ 0]}{\sigma_{\mathrm{eROSITA}, i}} \right)^2
    \label{eq:chi_squared}
\end{equation}
where $\phi_{\mathrm{DM},i}(p, m_\chi)$ is the DM or PBH induced X-ray flux in ring $i$ for $p = \{\langle\sigma v\rangle, \tau, f_{\mathrm{PBH}}\}$ for annihilation and decay of DM and PBH evaporation respectively, $\phi_{\mathrm{eROSITA}, i}$ is the measured eROSITA flux in ring $i$, and $\sigma_{\mathrm{eROSITA}, i}$ is the error in the eROSITA flux in ring $i$. We assume that the the measured counts in the eROSITA detector follow Poisson statistics, such that the error for $N$ counts  is equal to $\sqrt{N}$.  We vary $p$ to achieve $\chi^2=4$, representing a 2$\sigma$ confidence limit on $p$.

\begin{figure}
    \centering
    \includegraphics[width=0.49\linewidth]{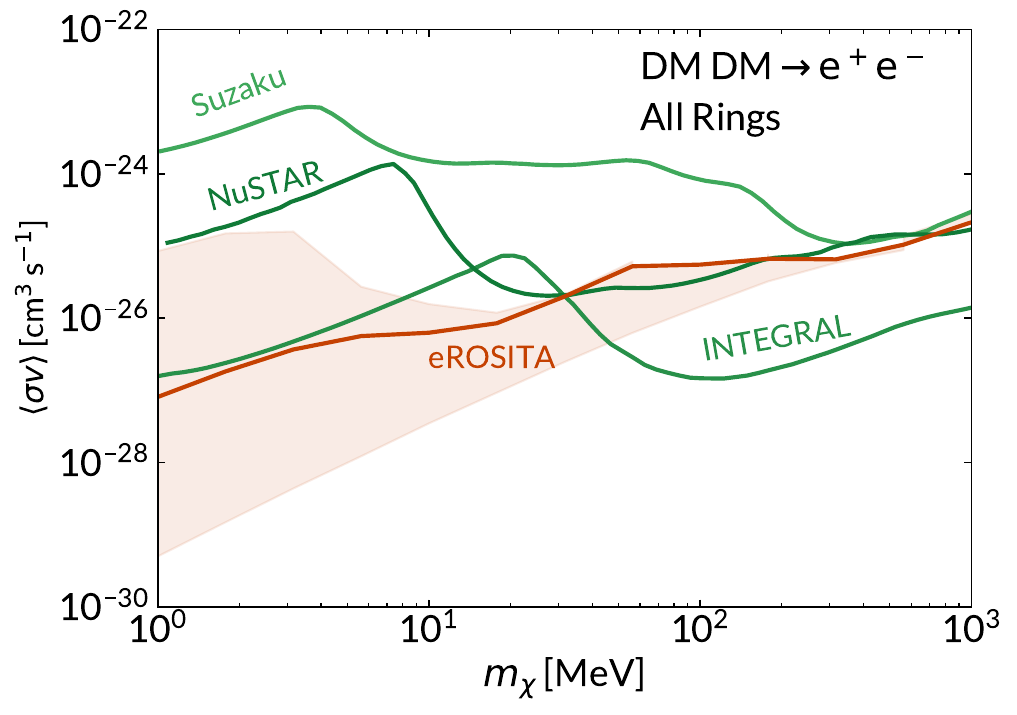}
\includegraphics[width=0.49\linewidth]{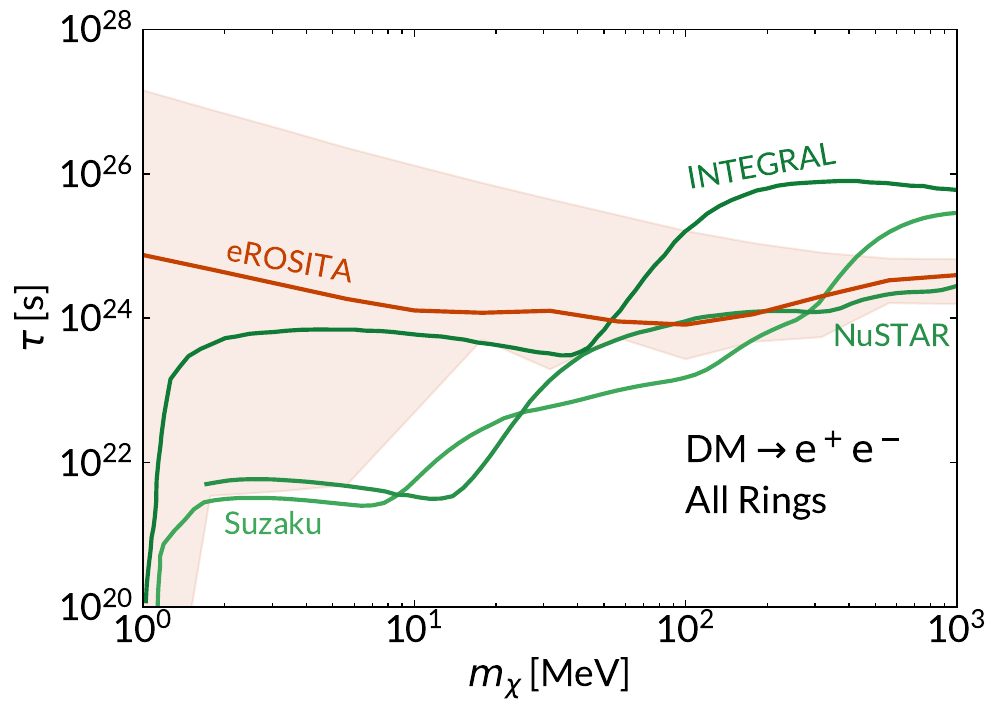}
\includegraphics[width=0.49\linewidth]{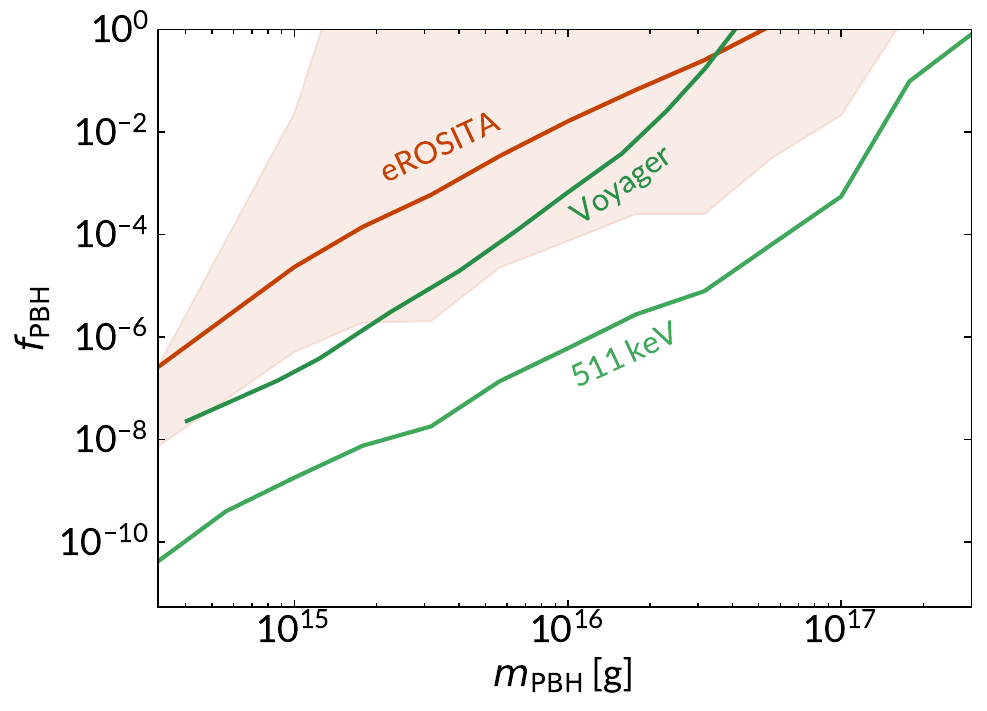}
\includegraphics[width=0.49\linewidth]{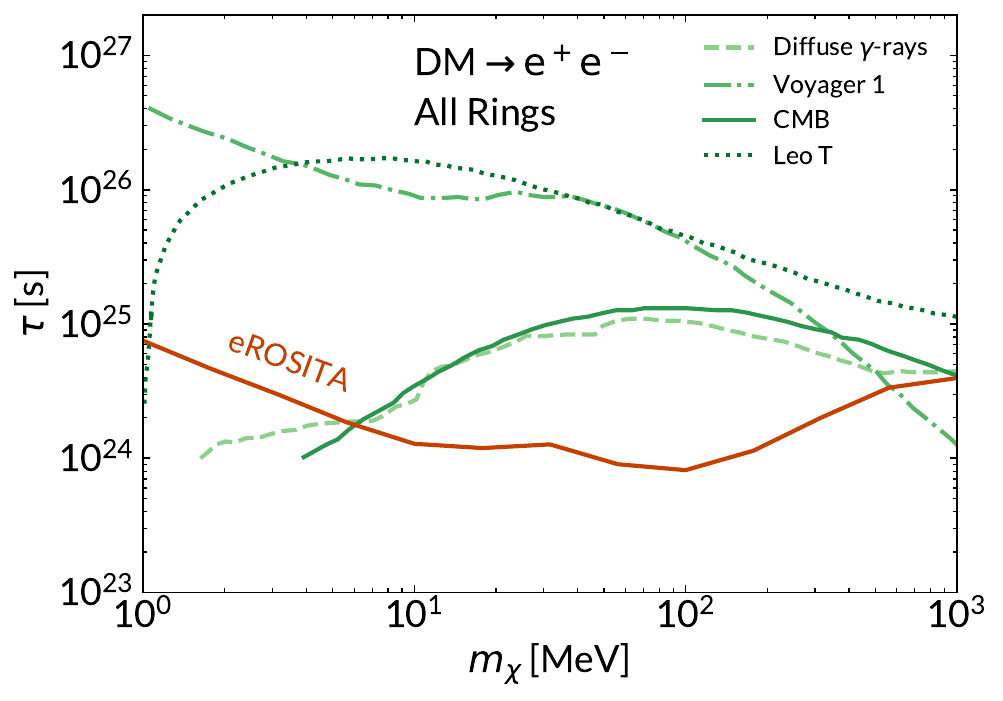}
    \caption{Comparison of the $95\%$ confidence bounds on annihilating (top left panel) and decaying (top right panel) DM with previous CR constraints. The bottom left panel shows constraints on the fraction of PBHs as DM derived in this work. Finally, in the bottom right panel, we show a comparison of the eROSITA decay constraints with other probes. In all the cases we report the results using the best-fit propagation parameters for eROSITA (orange line), as well as uncertainties arising from using pessimistic and optimistic propagation parameters (reddish band), alongside other existing constraints (green lines).}
    \label{fig:DM_constraints}
\end{figure}

To account for uncertainties in propagation, we evaluate constraints under three different propagation scenarios: standard, optimistic, and pessimistic (following Refs.~\cite{DelaTorreLuque:2023olp, DelaTorreLuque:2024qms}). These scenarios correspond to different values of key propagation parameters. Specifically, we vary the Alfvén velocity $v_A$ as 13.43, 40, and 0 km s$^{-1}$, and the halo height $H$ as 8, 16, and 4 kpc, for the standard, optimistic, and pessimistic cases, respectively. Since the diffusion coefficient $D$ is constrained by the ratio $D/H$, it is adjusted accordingly to maintain consistency with the secondary-to-primary CR ratios measured by AMS-02~\cite{AMS02, Derome_2019, Luque:2021nxb}.

In Fig.~\ref{fig:DM_constraints}, we show the constraints on the DM annihilation cross section, $\langle\sigma v\rangle$, and decay lifetime, $\tau$, into electrons and positrons, assuming a nominal NFW DM profile, using our combined ring analysis (orange line). The orange uncertainty bands represent those arising from varying of the propagation parameters in the scenarios described above. We also show existing constraints from INTEGRAL, NuSTAR and Suzaku (green lines)~\cite{Cirelli:2023tnx}. Compared to these probes, eROSITA provides the strongest constraints for DM masses up to a few tens of MeV. If we take constraints from rings individually, we find that rings 1 - 3 provide the strongest constraints, nearly identical to constraints derived from the full ring analysis for annihilation, as opposed to decay which benefits more from using all rings. As we observe, the uncertainties are larger for lower DM masses, which is due to the impact of reacceleration in the $e^{\pm}$ products: the lower their energy, the more prone they are to being boosted by the interaction with plasma waves~\cite{Strong_1998, seo1994stochastic}, as extensively reported and discussed in Refs.~\cite{DelaTorreLuque:2023olp, Boudaud:2016mos}.

To highlight the importance of these constraints, we show a comparison of the decay constraints obtained with eROSITA in comparison with those obtained from cosmological measurements, heating of dwarf galaxies and direct observations of CRs (in particular, from Voyager-1) and diffuse X- and $\gamma$-rays. We are not showing constraints from the 511 keV line~\cite{DelaTorreLuque:2023cef} in this plot, which can reach lifetimes of $10^{29}$~s for $1$~MeV DM particles. The cosmological constraints are from CMB measurements and are obtained from the study of anisotropies in the CMB power spectrum~\cite{Slatyer:2015jla, Lopez-Honorez:2013cua}, the constraint from Leo-T dwarf galaxy was derived requiring the rate of heat injection by DM to not exceed the ultra-low radiative cooling rate of gas observed~\cite{Wadekar:2021qae}, and those from diffuse $\gamma$-rays are constraints from the FSR emission associated with DM~\cite{Essig:2013goa} (obtained with COMPTEL, INTEGRAL and HEAO-I). Meanwhile, the constraint from Voyager-1 data is obtained by comparing the expected interstellar flux of electrons and positrons from DM with the flux measurements on board the probe. For the latter, it is important to note that in conservative conditions where reacceleration is very low, no limit can be obtained for DM masses below $\sim20$~MeV. As we see, the eROSITA constraints are especially competitive below $10$~MeV, although at higher masses they are weaker. However, we emphasise that these limits are quite conservative, given that we directly mask the inner $\pm 5^{\circ}$ around the Galactic plane\footnote{When doing the analysis adopting a $\pm 2^{\circ}$ mask, our constraints improve by no more than a factor of 2.}. On top of this, we note that including the astrophysical background components could lead to up to even an order of magnitude stronger constraints from eROSITA.

\begin{figure}
    \centering
    \includegraphics[width=0.5\linewidth]{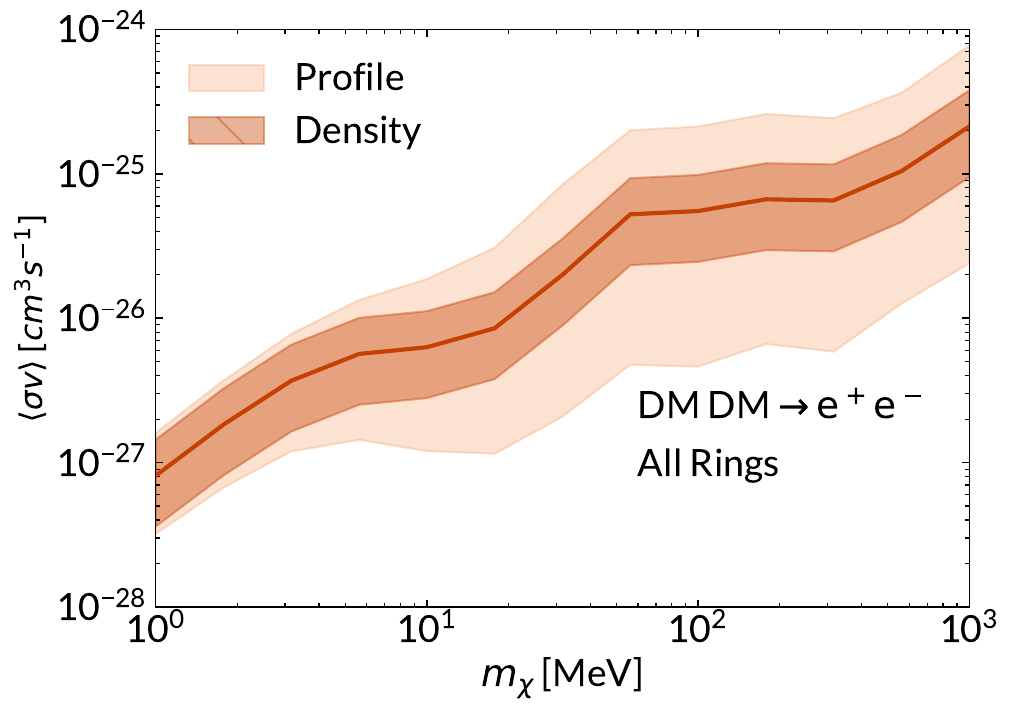}
\includegraphics[width=0.5\linewidth]{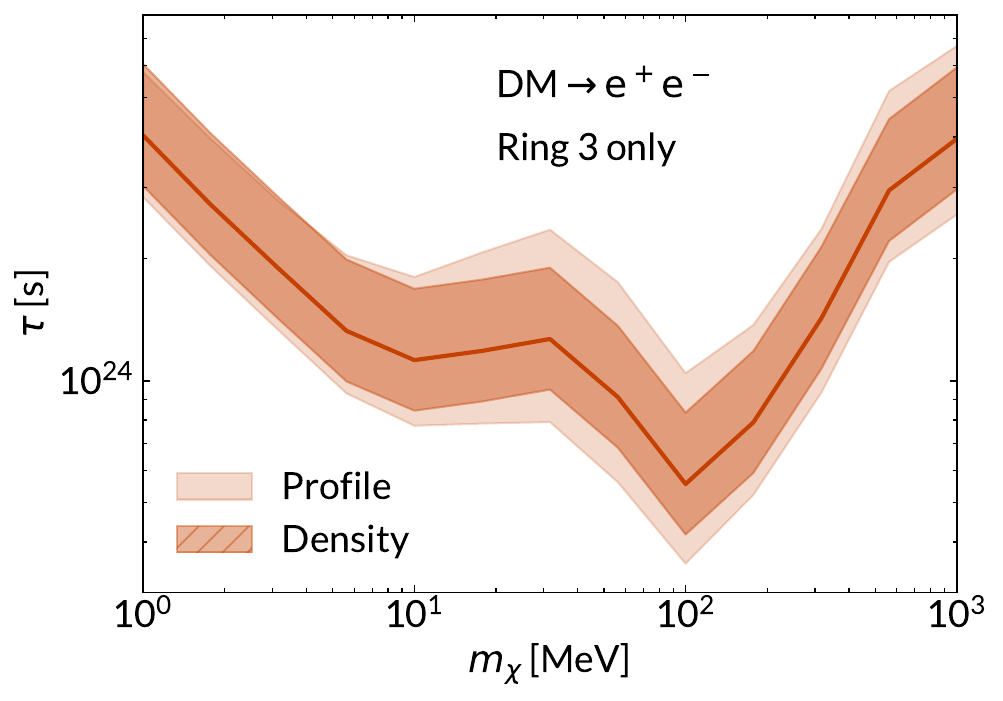}
    \caption{Comparison of the $95\%$ confidence bounds on annihilating (left panel) and decaying (right panel) DM with bands indicating the uncertainty from the DM density. The central line shows constraints from choosing the standard NFW profile and assuming a local DM density of $0.4$~GeV/cm$^3$. Shaded bands show uncertainty arising from considering a local DM density ranging from $0.3$ to $0.6$~GeV/cm$^3$ (darker bands) and varying the density profile from cNFW to isothermal distributions (lighter bands).}
    \label{fig:profile_uncertainty}
\end{figure}

We also quantify the uncertainty in the constraint arising from the DM density estimation. In particular, we show a band that represents the uncertainty due to the DM density profile adopted (Sec.~\ref{sec:Densities}) and another band that represents the uncertainty in determining the local DM density. To achieve this, we simulate the DM continuum emission for a contracted NFW (cNFW) profile with $\gamma = 1.4$ which has a steeper profile than the standard NFW, and an isothermal distribution, which is more shallow.  As discussed above, these represent two opposite and extremal scenarios that can significantly boost or suppress the diffuse X-ray emission generated by DM.

In these cases, we use the standard propagation parameters, as explained above. In Fig.~\ref{fig:profile_uncertainty}, we show the DM constraint for NFW (orange line), and the uncertainty derived from the adoption of cNFW and isothermal distributions (orange band) for both annihilation and decay scenarios. 
In the case of annihilation, the DM induced flux goes as $\phi_{\mathrm{DM}} \sim\rho_{\mathrm{DM}}^2$, making the profile uncertainties more dominant. Meanwhile, for decay, the flux dependence is weaker, varying as $ \phi_{\mathrm{DM}}\sim\rho_{\mathrm{DM}}$, and the uncertainty from the adopted profile is much lower, as expected. 
From this comparison we conclude that while above a few MeVs the DM density profile is the main source of uncertainty in these limits, below 10 MeV, propagation uncertainties dominate. 

Finally, in the bottom left panel of Fig.~\ref{fig:DM_constraints}, we also report constraints on PBHs in the mass range $10^{14} \ \mathrm{g} \lesssim m_{\mathrm{PBH}} \lesssim 10^{17} \ \mathrm{g} $ from Hawking evaporation of charged particles (orange line). As commented above, these constraints are derived assuming monochromatic and non-rotating PBHs. The orange band again represent uncertainties arising from propagation. 
Here, we also compare our result from eROSITA with other leading constraints, including the 511 keV line and Voyager 1 (as green lines). In comparison to these probes, eROSITA diffuse background observations do not provide competitive constraints.

\section{Revising {\sc XMM-Newton} constraints}
\label{sec:XMM}
The DM constraints derived from eROSITA data are significantly weaker than those reported in earlier analyses of {\sc XMM-Newton} observations~\cite{Cirelli:2023tnx, DelaTorreLuque:2023olp}. At first glance, this appears counterintuitive. While eROSITA probes lower energy bands than {\sc XMM-Newton}, where the latter suffers from increased instrumental noise, one would still expect comparable fluxes in the same regions of the sky when interpreted correctly. 

We believe this discrepancy arises from a misinterpretation in these previous analyses regarding the solid angle associated with the data presented in Ref.~\cite{XMM}. In particular, earlier works assumed the reported fluxes were distributed over the full geometric solid angle of each ring. However, the data in Ref.~\cite{XMM} are normalised to the exposure-weighted average solid angle $\Omega_{\mathrm{exp}}$, which accounts for the non-uniform exposure across the ring due to the telescope’s observational footprint\footnote{A Jupyter notebook tutorial on how to interpret the data is now provided in the original \href{https://github.com/bsafdi/XMM_BSO_DATA/blob/main/tutorial/Load_Data.ipynb}{Github repository}.}. This is defined
\begin{equation}
    \Omega_\mathrm{exp} = \int_\textrm{R} d\Omega \frac{T_{\mathrm{eff}}(\Omega)}{T_{\mathrm{tot}}(\Omega)},
\end{equation}
where $T_{\mathrm{eff}}$ is the exposure time in the angular direction and $T_{\mathrm{tot}}$ is the normalising total exposure time of the experiment, integrating over all angular directions in the ring region $\textrm{R}$. Crucially, these exposure-weighted solid angles are several orders of magnitude smaller than the corresponding geometric solid angles for each ring. For example, the geometric solid angle of ring 3 is $\Omega \approx 0.16 \ \mathrm{sr}$, whereas the exposure-weighted solid angle for {\sc XMM-Newton} in ring 3 is $\Omega_{\mathrm{exp}} \approx 3.5 \times 10^{-5} \ \mathrm{sr}$.

As a result, when DM-induced X-ray fluxes were computed assuming the geometric solid angle, the flux was overestimated by several orders of magnitude relative to the flux actually reported by {\sc XMM-Newton}, which properly accounts for the limited exposure. This can lead to spuriously strong DM constraints being derived.

\begin{figure}[!t]
    \centering
    \includegraphics[width=0.65\linewidth]{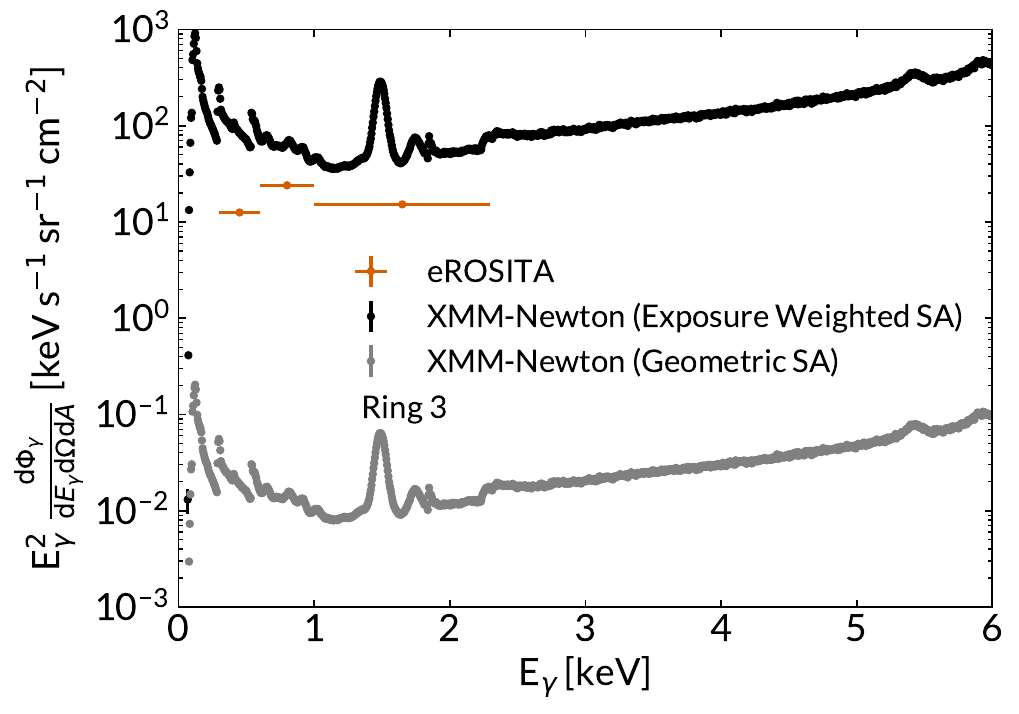}
      \caption{Comparison of ring 3 flux for eROSITA (orange points) and {\sc {\sc XMM-Newton}} MOS with the use of exposure-weighted solid angle (black points) consistent with observational coverage of the dataset and with direct use of the geometric solid angle (grey points). In order to enable a viable comparison, all fluxes shown here use a 2$^\circ$ mask of the Galactic plane, as provided for the {\sc {\sc XMM-Newton}} data.}
    \label{fig:XMM_eROSITA_comparison}
\end{figure}

To emphasise this, in Fig.~\ref{fig:XMM_eROSITA_comparison} we plot the ring 3 fluxes (using a comparable $|b| < 2^\circ$ inner mask of the Galactic plane) recorded for both eROSITA (orange points) and {\sc XMM-Newton} assuming the exposure-weighted average solid angle (black points) as intended in the data analysis and the geometric solid angle (grey points) in the energy range $0$-$6$ keV. We use experiment independent units of $\mathrm{keV\ s^{-1} \ sr^{-1} \ cm^{-2}}$ to produce compatible spectra. We see that using the geometric solid angle significantly underestimates the true flux and, in fact, {\sc XMM-Newton} should provide somewhat weaker or comparable constraints to eROSITA once this has been addressed. 

\begin{figure}[!ht]
    \centering
\includegraphics[width=0.49\linewidth]{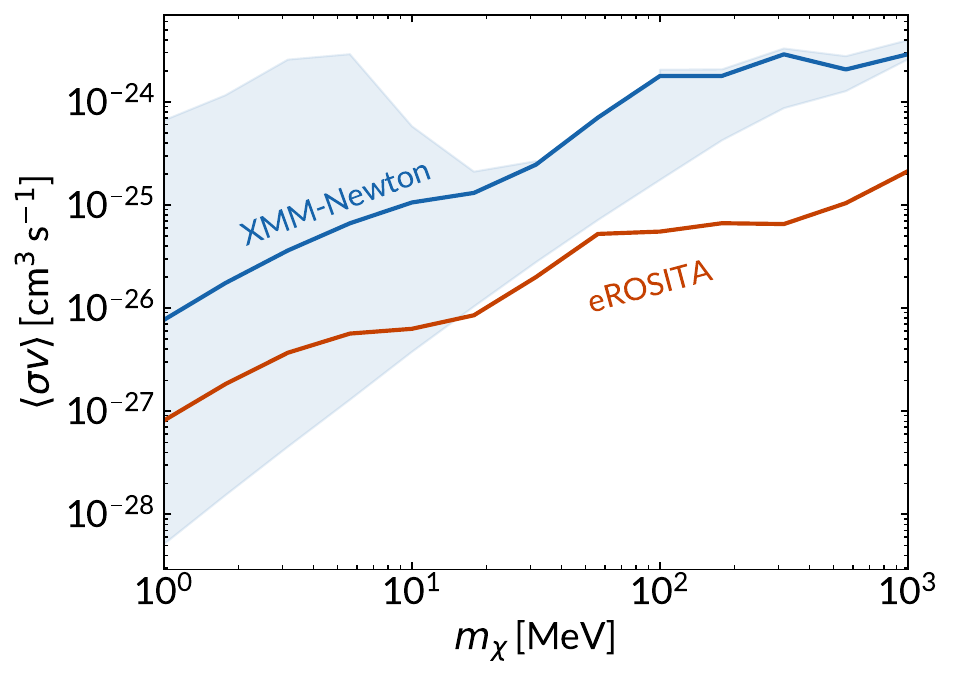}
\includegraphics[width=0.49\linewidth]{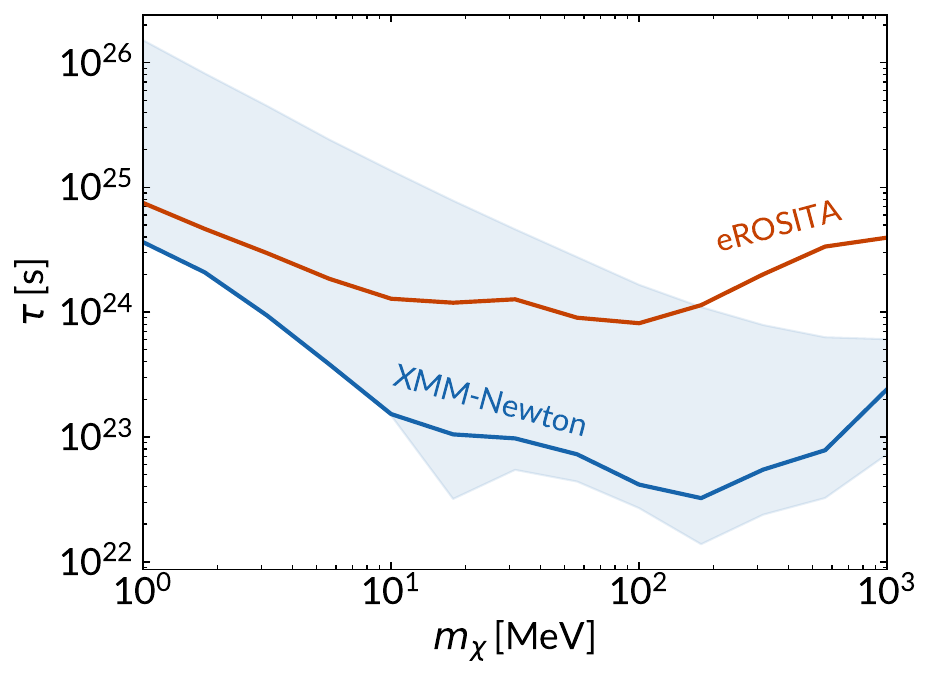}
\includegraphics[width=0.49\linewidth]{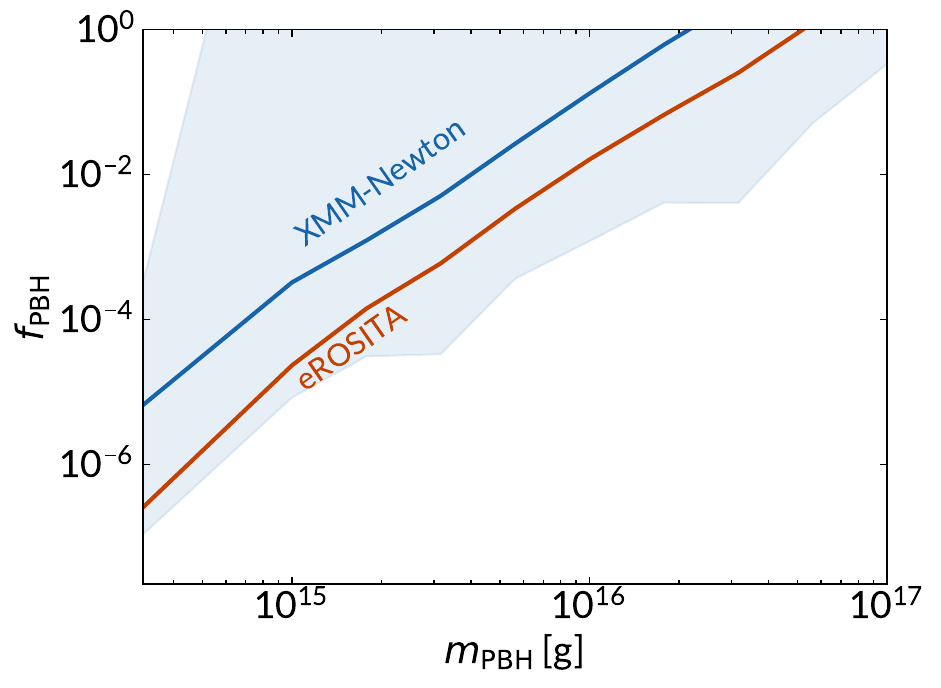}
    \caption{Comparison of the $95\%$ confidence bounds on annihilating (top left panel) and decaying (top right panel) sub-GeV DM and evaporating PBHs (bottom panel) between eROSITA (red) and {\sc XMM-Newton} (blue) using the exposure-weighted solid angle for the best-fit propagation parameters. In the case of {\sc XMM-Newton}, we show uncertainties arising from propagation parameters (blue shaded region).}
    \label{fig:XMM_new_constraints}
\end{figure}

The discrepancy in solid angle varies between rings, so a full reanalysis is required to derive revised {\sc XMM-Newton} constraints. Using the same analysis framework as for eROSITA, we compute updated {\sc XMM-Newton} limits on DM annihilation and decay into $e^\pm$, as well as PBH evaporation, shown in Fig.~\ref{fig:XMM_new_constraints}. Here we consider energies from 2.5-8 keV, identical to previous analyses. These incorporate propagation uncertainties consistent with those used in the eROSITA analysis. We do not repeat the uncertainty propagation for the DM density distribution or the local DM density as this can be inferred easily from Fig.~\ref{fig:profile_uncertainty}.

For annihilation, the revised {\sc XMM-Newton} bounds are roughly an order of magnitude weaker than those from eROSITA across the full DM mass range. For decaying DM, {\sc XMM-Newton} constraints are more competitive at low masses, where FSR dominates, but become nearly two orders of magnitude weaker at higher masses, where IC emission governs the size of the signal. For PBHs, we find that {\sc XMM-Newton} constraints are consistently about an order of magnitude weaker than those from eROSITA over the full PBH mass range considered. This clarification realigns the role of {\sc XMM-Newton} in the broader landscape of X-ray DM searches and confirms that eROSITA currently provides the strongest diffuse Galactic X-ray limits in the sub‑10 MeV regime, though other probes such as Voyager and 511 keV line studies remain more constraining in certain regions of parameter space.

\section{Discussion and conclusions}
\label{sec:Conclusion}

X-ray observations of diffuse Galactic emission offer a powerful avenue to probe new physics scenarios. In particular, sub-GeV DM particles that annihilate or decay into electron-positron pairs populate the interstellar medium with low-energy charged particles. These electrons and positrons interact with Galactic ISRFs via IC scattering and bremsstrahlung, generating diffuse X-ray emission peaking around the keV scale. This makes eROSITA an especially sensitive probe for such signatures, particularly since IC scattering of MeV-scale electrons on the CMB peaks just below 1 keV. Similarly, sub-asteroid mass PBHs emit low-energy $e^\pm$ via Hawking evaporation, leading to analogous X-ray signals.

In this work, we used the first all-sky data release from eROSITA (eRASS1) to derive new constraints on sub-GeV DM and asteroid-mass PBHs. We constructed sky maps of the diffuse X-ray background in the 0.3–2.3 keV range and divided the sky into 30 concentric rings centred on the GC, excluding the central $|b| < 5^\circ$ region to minimise absorption from Galactic gas. These regions were then combined to place 95\% confidence level limits.

We find that, even with just six months of data, eROSITA provides competitive X-ray constraints. For decaying DM below $\sim 10$ MeV, our limits surpass those from CMB and diffuse $\gamma$-ray observations, though they remain weaker than bounds from Voyager and Leo‑T dwarf galaxy heating in the same mass range. However, in this low-mass regime, signal predictions are subject to larger uncertainties due to charged particle propagation modeling. Looking ahead, eROSITA’s ability to resolve point sources will improve with future data releases, allowing for better subtraction of unresolved backgrounds and potentially yielding significantly stronger constraints. Current diffuse emission is dominated by unresolved sources such as coronal stellar emission and accreting white dwarfs. Improved estimates of this unresolved component could tighten the bounds presented here by orders of magnitude.

We also note that our analysis conservatively excluded the Galactic plane, where DM density is highest. Future work incorporating detailed modeling of X-ray absorption by interstellar gas could improve sensitivity, although CMB constraints will likely remain stronger for annihilating DM.

Finally, we reanalysed previous constraints from {\sc XMM-Newton} data in Ref.~\cite{Cirelli:2023tnx} and subsequently in Ref.~\cite{DelaTorreLuque:2023olp}, where we identified a discrepancy arising from the treatment of the instrument’s exposure-weighted solid angle. Specifically, an overestimate of the effective sky coverage led to flux values, and therefore constraints, being overstated by approximately four orders of magnitude. After correcting this effect and rederiving the {\sc XMM-Newton} limits using consistent assumptions, we find that they are weaker than those derived from eROSITA.

In summary, eROSITA provides the strongest diffuse X-ray limits from annihilating and decaying dark matter in the $\lesssim10$ MeV mass range, surpassing previous X-ray bounds from INTEGRAL, NuSTAR, and Suzaku. For decaying dark matter, our constraints also improve upon those from the CMB and diffuse $\gamma$-ray observations, though they remain weaker than the most stringent bounds from Voyager-1 and Leo-T dwarf galaxy heating. For PBHs, eROSITA limits are weaker than those from Voyager and 511 keV studies. Nevertheless, its diffuse all-sky coverage makes it uniquely suited to provide complementary and robust checks, especially with future deeper data releases.

\section*{Acknowledgements}
We thank Joshua Foster for providing useful input regarding the exposure-weighted solid angle for {\sc XMM-Newton} data analysis. PDL thanks Jari Kajava and Jan-Uwe Ness for their feedback and useful discussions on the topic.
SB is supported by the STFC under grant ST/X000753/1.
PDL is supported by the Juan de la Cierva JDC2022-048916-I grant, funded by MCIU/AEI/10.13039/501100011033 European Union ``NextGenerationEU"/PRTR. The work of PDL is also supported by the grants PID2021-125331NB-I00 and CEX2020-001007-S, wich are both funded ``ERDF A way of making Europe'' and by MCIN/AEI/10.13039/5011000\\11033. PDL also acknowledges the MultiDark Network, ref. RED2022-134411-T. This project used computing resources from the Swedish National Infrastructure for Computing (SNIC) under project Nos. 2021/3-42, 2021/6-326, 2021-1-24 and 2022/3-27 partially funded by the Swedish Research Council through grant no. 2018-05973. DC acknowledges support
from a Science and Technology Facilities Council (STFC) Doctoral Training Grant. This project used computing resources from King's College London CREATE system \cite{KCLCREATE2025}. For the purpose of open access, the authors have applied a Creative Commons Attribution (CC
BY) license to any Author Accepted Manuscript version
arising from this submission. 


\bibliographystyle{JHEP}
\bibliography{references}

\appendix
\section{Appendix: eROSITA constraints for $\mu^+\mu^-$ and $\pi^+\pi^-$ channels.}
\begin{figure}
    \centering
    \includegraphics[width=0.48\linewidth]{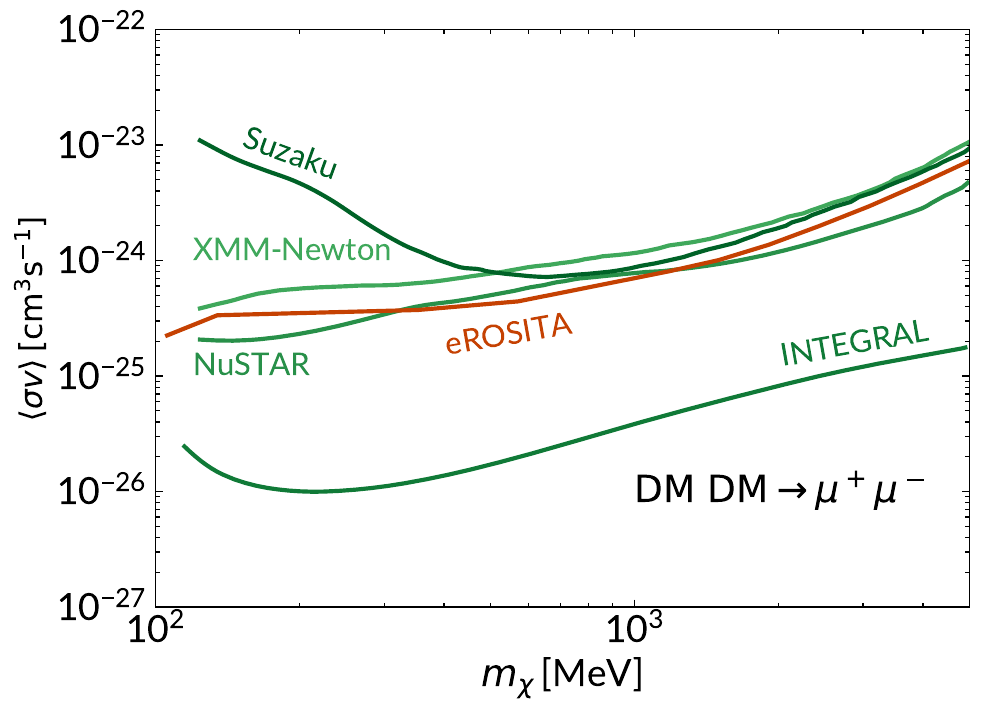}
    \includegraphics[width=0.48\linewidth]{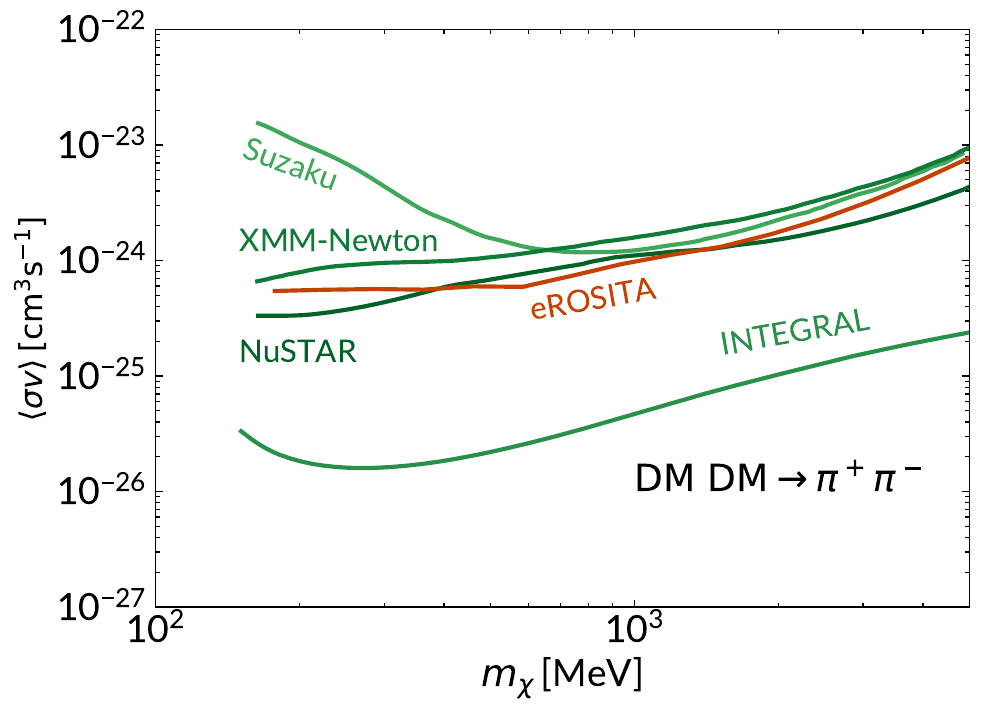}
    \includegraphics[width=0.48\linewidth]{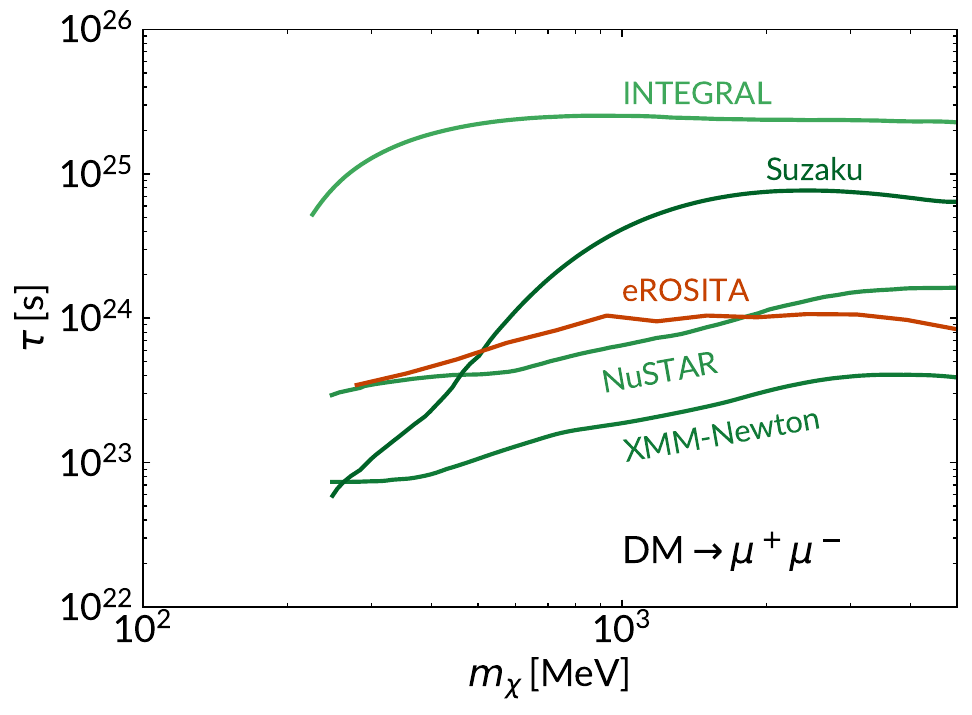}
    \includegraphics[width=0.48\linewidth]{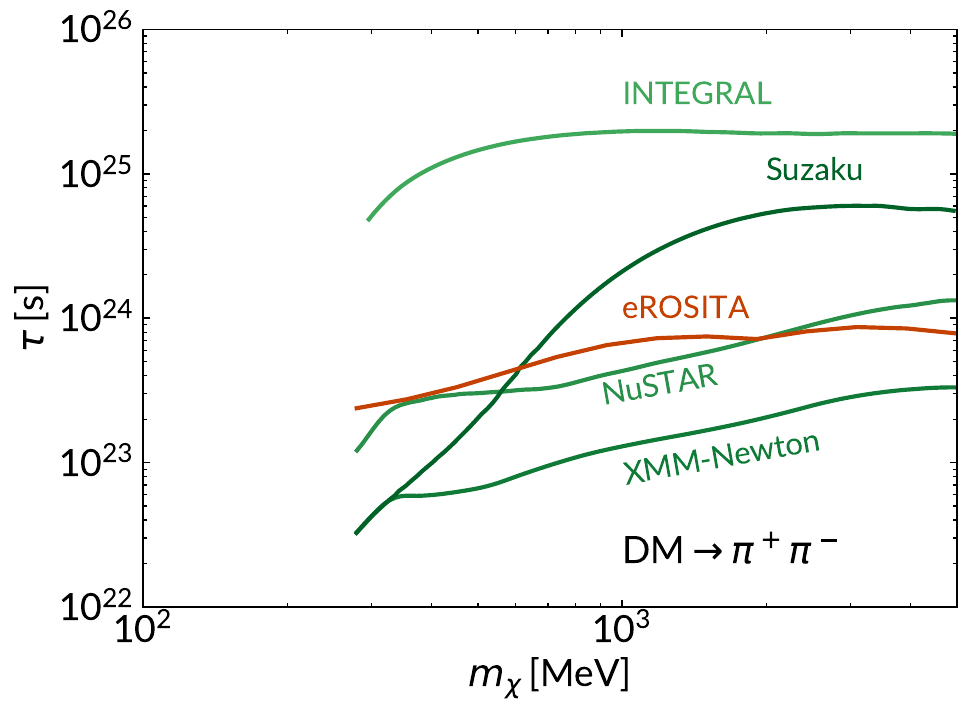}
    \caption{Comparison of the 95\% confidence bounds on DM annihilation cross-section (top panels) and decay lifetimes (bottom panels) into $\mu^+\mu^-$ (left panels) and $\pi^+\pi^-$ (right panels) from eROSITA (orange line) and the constraints from INTEGRAL, Suzaku, NuSTAR and revised XMM-Newton constraints \cite{Cirelli:2023tnx} (green lines).}
    \label{fig:pipi_mumu_constraints}
\end{figure}

In this appendix we derive the constraints for DM annihilation and decay channels into $\mu^+\mu^-$ and $\pi^+\pi^-$ channels which subsequently produce $e^+e^-$ as final products. The only computational difference is replacing the electron injection spectra $dN_e/dE_e$ in Eq.~\eqref{eq:Source} for each of the channels. We use the injection spectra computed in Ref.~\cite{Cirelli:2020bpc}. The rest of the procedure is identical to the direct $e^+e^-$ case. For simplicity, we restrict ourselves to the standard propagation parameter setup, using a NFW DM profile. The resulting limits for eROSITA can be found in Fig.~\ref{fig:pipi_mumu_constraints}, alongside constraints derived from other X-ray probes in Ref.~\cite{Cirelli:2023tnx}\footnote{An erratam of this reference was published onto the arXiv on the 17th of July 2025 following our initial publication of this work, which has updated the XMM-Newton limits based on the correction we outlined in Section~\ref{sec:XMM}.}. 

We see that eROSITA provides comparible constraining power to NuSTAR, XMM-Newton and Suzaku probes in the case of annihilation into these channels. For decay, eROSITA provides moderate constraints compared to these probes. However, for each channel considered, INTEGRAL constraints are strongest across the whole mass range. This is not surprising, as the shape of the primary electron spectra for these channels is comparable to direct $e^+e^-$ \cite{DelaTorreLuque:2023olp}, therefore producing similar IC X-ray spectra. eROSITA loses sensitivity to INTEGRAL for direct $e^+e^-$ production for dark matter masses above $\sim 100 \ \mathrm{MeV}$ (see Fig.~\ref{fig:DM_constraints}), and this is the only region which is kinematically accessible to these heavier channels, and therefore the constraining power of eROSITA is of lesser interest for $\mu^+\mu^-$ and $\pi^+\pi^-$ channels.

\end{document}